\documentclass[runningheads]{llncs}
\usepackage[l2tabu,orthodox]{nag}
\usepackage[utf8]{inputenc}
\usepackage[british]{babel}
\usepackage[caption=false]{subfig}
\usepackage{balance}
\usepackage{booktabs}
\usepackage{upquote}
\usepackage{graphicx}
\usepackage{url}
\usepackage{tabularx}
\usepackage{ltablex}

\usepackage{listings}
\newcommand{\pb}[1]{\noindent{\textbf{\emph{#1}}.}}

\usepackage{xcolor}

\usepackage{orcidlink}

\newcommand{\NumberOfProbesThatDidThreeTests}{6154}
\newcommand{\NumberOfProbesThatDidCustomPingTest}{765}
\newcommand{\ConclusivelyPassedAnyTests}{255}
\newcommand{\PassedDNSOneAndTwo}{36}
\newcommand{\SizeNATSixFourPlusDNSSixFour}{18}
\newcommand{\NATSixFourOnlyCustomPing}{160}
\newcommand{\NATSixFourOnlyStdPing}{52}
\newcommand{\NoConfiguredNAT}{139}
\newcommand{\UsesPublicResolver}{56}
\newcommand{\PassedDNSOneNotPing}{162}
\newcommand{\FreeASOnlyPassedDNSOne}{152}
\newcommand{\TotalDualStack}{205}
\newcommand{\TotalNumNATSixFourProbes}{224}
\newcommand{\TotalNumvFourASN}{143}
\newcommand{\TotalNumvSixASN}{43}
\newcommand{\FailedDNSOneTwoPassedPingTest}{326}
\newcommand{\SizeNATSixFourOnly}{206}

\newcommand{\TotalNumberProbesRedoTraceroute}{183}
\newcommand{\TotalNumberPathsRedoTraceroute}{3565}
\newcommand{\NumberOfPathsNotVisNat}{2320}
\newcommand{\TotalNumberPathsVisNatTraceroutes}{1230}
\newcommand{\TotalNumberProbesVisNatTraceroute}{54}
\newcommand{\TotalNumberProbesRecurringInitial}{37}
\newcommand{\TotalNumberPathsRecurringRemoved}{1076}
\newcommand{\TotalNumberProbesRecurring}{34}
\newcommand{\RecurringNATSixFourOnly}{28}
\newcommand{\RecurringNATSixFourPlusDNSSixFour}{6}
\newcommand{\OnePrefixProbesRecurring}{27}
\newcommand{\IPvFourTotalSuccessRate}{87.94}
\newcommand{\IPvSixTotalSuccessRate}{87.07}
\newcommand{\DiffIPvFourIPvSixSuccessRate}{0.87}
\newcommand{\IPvFourReachedASRate}{12.03}
\newcommand{\IPvSixReachedASRate}{0.00}
\newcommand{\NumPathsBothSuccessful}{13519}
\newcommand{\PercentBothPathsSuccessful}{86.86}

\newcommand{\NumberNATSixFourOnlyLocalNAT}{23}
\newcommand{\MeanIPvFourPathLength}{17.72}
\newcommand{\StdDevIPvFourPathLength}{5.25}
\newcommand{\MedianIPvFourPathLength}{17.00}
\newcommand{\MeanIPvSixPathLength}{20.99}
\newcommand{\StdDevIPvSixPathLength}{6.10}
\newcommand{\MedianIPvSixPathLength}{21.00}
\newcommand{\AvgPathLenDiff}{3.27}
\newcommand{\PercentLenIncrease}{23.13}
\newcommand{\PercentRTTIncrease}{17.47}
\newcommand{\LocalNATMeanPathLenDiff}{3.45}
\newcommand{\RemoteNATMeanPathLenDiff}{2.21}
\newcommand{\RemoteNATMeanPathLenDiffNoOutlier}{5.08}
\newcommand{\LocalNATMeanRTTDiff}{27.56}
\newcommand{\RemoteNATMeanRTTDiff}{1.22}
\newcommand{\AvgIPvFourRTT}{160.82}
\newcommand{\StdIPvFourRTT}{91.61}
\newcommand{\MedianIPvFourRTT}{169.80}
\newcommand{\AvgIPvSixRTT}{184.55}
\newcommand{\StdIPvSixRTT}{105.59}
\newcommand{\MedianIPvSixRTT}{193.51}
\newcommand{\AvgRTTDiff}{23.74}
\newcommand{\LocalRTTDiffNoXLAT}{1.34}
\newcommand{\CorrelationCoefficientLenDiffRTTDiff}{0.29}
\newcommand{\AvgMissingHopsIPvFour}{16.52}
\newcommand{\StdMissingHopsIPvFour}{11.92}
\newcommand{\MedianMissingHopsIPvFour}{16.67}
\newcommand{\AvgMissingHopsIPvSix}{38.06}
\newcommand{\StdMissingHopsIPvSix}{14.50}
\newcommand{\MedianMissingHopsIPvSix}{36.84}
\newcommand{\MeanSameMissing}{89.50}
\newcommand{\MedianSameMissing}{100.00}
\newcommand{\StdSameMissing}{16.46}
\newcommand{\XLATAverageLenDiff}{4.38}
\newcommand{\NoXLATAverageLenDiff}{2.49}
\newcommand{\TotalNumberPathsRecurring}{15564}
\newcommand{\TotalProbesLocalNAT}{27}

\begin{document}

\title{Measuring NAT64 Usage in the Wild}

\author{Elizabeth Boswell\inst{1}\orcidlink{0000-0003-2388-3806} \and 
        Stephen McQuistin\inst{2}\orcidlink{0000-0002-0616-2532} \and
        Colin Perkins\inst{3}\orcidlink{0000-0002-3404-8964} \and
        Stephen Strowes\inst{4}\orcidlink{0000-0002-6341-3406}}
\authorrunning{E. Boswell et al.}
%
\institute{University of Glasgow
\email{e.boswell.2@research.gla.ac.uk} \and
University of St Andrews 
\email{sm@smcquistin.uk} \and
University of Glasgow
\email{csp@csperkins.org} \and
Fastly
\email{sds@fastly.com}\\
}

\maketitle


\begin{abstract}
  NAT64 is an IPv6 transition mechanism that enables IPv6-only hosts to access
  the IPv4 Internet. Understanding the deployment of NAT64, and its performance impact, is
  crucial to the success of the IPv6 transition, by encouraging IPv6-only
  deployments. We develop a set of tests for detecting
  NAT64 and apply them to the RIPE Atlas testbed, finding \TotalNumNATSixFourProbes\ probes,
  in  \TotalNumvSixASN\ networks, that can use NAT64 to access the IPv4 Internet.
  Using \TotalNumberProbesRecurring\ dual stack probes, that have both NAT64
  and native IPv4 access, to compare performance, we find that NAT64 paths
  are, on average, \PercentLenIncrease\% longer, with \PercentRTTIncrease\%
  higher round-trip times.

\end{abstract}

\section{Introduction}
With the IPv4 address space being depleted~\cite{richter_primer_2015},
and IPv6 adoption continuing to grow~\cite{google_v6_stats}, there is an
increased need for transition mechanisms to allow IPv4 and IPv6 hosts
to seamlessly communicate. Understanding the impact of these transition
mechanisms is important, both to support short-term IPv6 adoption and to
encourage longer-term IPv6-only deployments.

One transition mechanism is NAT64~\cite{matthews_stateful_2011}. Together
with DNS64~\cite{matthews_dns64_2011}, it defines a way to embed IPv4 addresses
within IPv6 addresses, enabling network address translation between IPv6 and
IPv4 so that IPv6-only devices can access the IPv4-only Internet.
Successful NAT64 deployments must be located so they don't become performance
bottlenecks or lead to paths that are longer, and
have higher latency, than native IPv4 paths. If NAT64 significantly impacts
performance, this will discourage IPv6-only deployments, slowing adoption.

In this paper, we conduct a preliminary study of NAT64 devices in the Internet.
We develop a set of tests to detect NAT64 and apply them on RIPE Atlas to
find probes that can use NAT64 to access the IPv4 Internet. We characterise
the location of these probes and their NAT64 devices, and compare
characteristics of NAT64-based and native IPv4 paths.

While several studies have explored NAT64 performance using small test networks
(e.g., \cite{llanto_performance_2012}, \cite{lencse_performance_2013},
\cite{tsetse_measuring_2012}), ours is one of the first large-scale studies of the
behaviour of NAT64 deployment in the wild, alongside \cite{hsuFirstLookNAT642024}\footnote{This work was not yet available to us at the time of writing.}. Specifically, we make the following
contributions:
\begin{itemize}
\item We develop a set of tests for detecting NAT64 (\S\ref{sec:nat64-search});

\item We apply those tests on RIPE Atlas to detect RIPE Atlas probes situated behind NAT64
      middleboxes, finding \TotalNumNATSixFourProbes\ such probes, in
		\TotalNumvSixASN\ networks (\S\ref{sec:nat64-ripe}), and categorise the types of deployments (\S\ref{sec:probe-categorisation}).

\item We compare IPv4 and NAT64 translated paths, finding that,
      on average, NAT64 paths are \PercentLenIncrease\% longer with
      \PercentRTTIncrease\% higher round-trip times (\S\ref{sec:traceroutes}).

\end{itemize}
By studying NAT64 performance in the wild, we demonstrate
some limitations of current deployments and provide a basis for further
study. 

\section{Motivation}
\label{sec:motivation}

As the Internet transitions from IPv4 to IPv6, a number of mechanisms have
been defined to allow IPv4 and IPv6 hosts to communicate~\cite{richter_primer_2015}.
These include NAT64~\cite{matthews_stateful_2011}, a network address
translation mechanism that allows IPv6-only hosts to access the IPv4
Internet, and the associated extensions to DNS, known as DNS64, that
synthesise \texttt{AAAA} records for names representing IPv4-only hosts.

\begin{figure}[t]
\centering
\includegraphics[width=0.7\columnwidth,trim={0 16mm 0 5mm}]{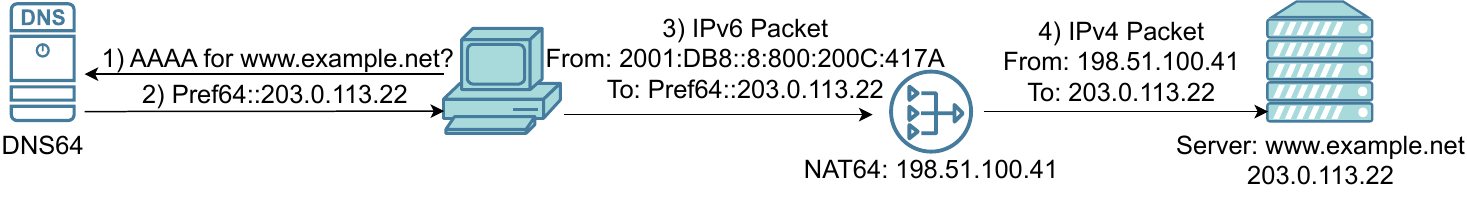}
  \caption{Address translation using DNS64 and NAT64.}
\label{fig:nat64-diagram}
\end{figure}

Figure~\ref{fig:nat64-diagram} shows how address translation is performed
using DNS64 and NAT64. An IPv6 host sends a DNS \texttt{AAAA} query
to the DNS64 resolver, for a domain name that only has an IPv4 address. The
DNS64 resolver synthesises a \texttt{AAAA} record containing an IPv6
address that encodes the IPv4 address of the target host. The IPv6 address
is either made up of the standard NAT64 prefix, \texttt{64:ff9b::/96}, or a
different custom prefix, followed by the IPv4 address, usually in the least
significant 32 bits of the IPv6 address~\cite{li_ipv6_2010}.

Packets sent to that synthesised address by the IPv6 host are routed to
the NAT64, which extracts the IPv4 address, translates the packet to IPv4,
and sends it on to the IPv4 host. Replies from the IPv4 host are similarly
translated back into IPv6. NAT64 devices can be set up by ISPs~\cite{chen_nat64_2014},
public providers (e.g., as listed at \url{https://nat64.xyz}), or
individual users~\cite{ripe72_hendriks}. 

Several small-scale studies have measured the performance
impact of NAT64 \cite{lencse_performance_2013,tsetse_measuring_2012,llanto_performance_2012}.
These characterise the translation
overhead, and the performance of different types of NAT, but in a testbed
environment. Accordingly, they do not necessarily reflect real-world use
in the Internet. Like~\cite{hsuFirstLookNAT642024}, we study the behaviour of
NAT64 in the wild, with comparative evaluation using the RIPE Atlas
platform. 

In the following, we first develop a set of tests that allow us to detect
the presence of a NAT64 on the network path. We apply these
tests to over 6,400 RIPE Atlas probes, to identify probes that
are using NAT64, potentially in combination with DNS64, and characterise these deployments. Finally, we run
UDP \texttt{traceroute} from these probes to measure the
path length and latency impact of the NAT64 deployments.

While \texttt{traceroute} measurements do not directly reflect the
performance of real-world applications, they can reveal some impediments to traffic.
Further, \texttt{traceroute} provides a fine-grained view of variation in 
path characteristics such as latency and number of hops. 
This allows us to measure the relative impact that the introduction of NAT64
has on these characteristics. Further application performance
issues are out of scope for this paper.

\pb{Relation to 464XLAT}
DNS64 and NAT64 assist the IPv6 transition in the common case where an
application resolves a service name via a \texttt{AAAA} DNS lookup, but
the service they are resolving has only published an \texttt{A} record.
They do not address cases where the host sends directly to an IPv4 address
without a DNS lookup (e.g., if it received the address via some out-of-band
means).

464XLAT~\cite{mawatari_464xlat_2013} is an IPv6 transition mechanism
that is intended to provide compatibility for such cases.
It specifies use of a customer-side NAT46 (``CLAT'') for
IPv4-to-IPv6 translation in addition to the provider-side NAT64.
464XLAT thus translates IPv4 packets twice: the CLAT translates
them from IPv4 to IPv6 on the host, for transit over the provider
network, then the provider-side NAT64 translates them back to IPv4 for onwards
delivery to their destination.
DNS64 is not strictly required in a 464XLAT deployment, but it is typical:
IPv6 address synthesis will reduce the number of address
family translations required to support application traffic.
Thus, DNS64 and NAT64 may be standalone from a 464XLAT environment, or may be
components of a 464XLAT deployment.
464XLAT is most typically deployed in cellular networks, but guidelines for
deployment in other network types are available~\cite{rfc8683}.

As we discuss in \S\ref{sec:nat64-search}, it is possible to
systematically discover DNS64 and NAT64 deployments. We have no equivalent
mechanism to detect the existence of a CLAT since the translation provided
by the CLAT hides IPv6 addresses on the path.
We therefore limit our results to measuring NAT64 and DNS64 deployment, and
do not further consider 464XLAT.
RIPE Atlas probes do not implement an on-device CLAT, though it may be
possible that a CLAT is implemented elsewhere in a network~\cite{rfc8683}.

\pb{Applicability}
We use RIPE Atlas as a measurement platform, giving wide coverage in
residential and operator networks but limited visibility into mobile
networks. The use of a wireless testbed platform, such as was previously
provided by MONROE, would give broader visibility into such networks but,
to the best of our knowledge, no such testbed is available at the time of writing.

\section{Detecting NAT64 Devices}
\label{sec:nat64-search}

\begin{figure}[t]
\centering
\includegraphics[width=0.75\columnwidth,trim={0 5mm 0 9mm}]{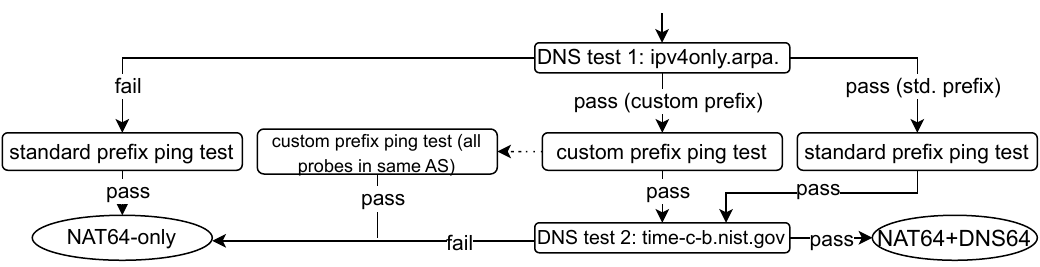}
\caption{NAT64 detection tests and result groups.}
\label{fig:nat64-search-procedure}
\end{figure}

We develop and use four tests to detect and characterise NAT64 devices: two tests of
DNS behaviour and two \texttt{ping} tests, described in
Figure~\ref{fig:nat64-search-procedure} and below.
These categorise probes into two groups: (i) those with a working NAT64 and
DNS64 resolver (NAT64+DNS64); and (ii) those with working NAT64 that lack access to a
DNS64 resolver to synthesise \texttt{AAAA} records for IPv4-only names (NAT64-only).
The four tests are as follows:
\begin{itemize}
  \item \textbf{DNS Test 1} uses the NAT64 prefix discovery procedure described
    in RFC 7050~\cite{savolainen_discovery_2013}. Hosts send a DNS
    \texttt{AAAA} query for the \texttt{ipv4only.arpa.} name. This is a
    special-use domain name that only resolves to IPv4, so where DNS64 is
    in use hosts will receive a synthetic \texttt{AAAA} record in response.
    IPv6 only hosts without DNS64 will receive \texttt{NXDOMAIN}.

  \item \textbf{DNS Test 2} is similar to DNS Test 1, but with a request for a
    different IPv4-only name (\texttt{time-c-b.nist.gov.}). This test is
    needed because we found that some hosts have access to a DNS64 resolver
    that can correctly synthesise a \texttt{AAAA} record for \texttt{ipv4only.arpa.},
    and hence pass DNS Test 1, but that fail to synthesise \texttt{AAAA}
    records for other IPv4-only names.

  \item The \textbf{Standard Prefix \texttt{ping} Test} involves sending
    \texttt{ping} requests to the address of a widely available IPv4-only host
    (\texttt{91.201.7.243}; RIPE Atlas anchor probe \#6771), encoded into IPv6
    as \texttt{64:ff9b::5bc9:7f3} using the standard NAT64 prefix. This test is
    used to confirm that hosts receiving the standard prefix from DNS64 can
    \texttt{ping} addresses in that prefix. 

  \item The \textbf{Custom Prefix \texttt{ping} Test} is used when DNS Test 1
    returns a \texttt{AAAA} record using a non-standard NAT64 prefix (i.e.,
    not \texttt{64:ff9b::/96}). The custom prefix is used to encode the same
    IPv4 address as in the Standard Prefix ping Test (i.e., \texttt{91.201.7.243}),
    which is then used as the target of an IPv6 \texttt{ping} request to
    confirm that the NAT64 works. 

\end{itemize}

All probes perform the Standard Prefix \texttt{ping} Test, regardless of whether they pass any DNS tests. This allows us
to find probes that can \texttt{ping} standard
prefix addresses but didn't receive the prefix from DNS64. 
The Custom Prefix \texttt{ping} Test is applied to probes
within the same IPv6 AS as a probe that received a non-standard prefix.
These additional tests detect probes that use a DNS
resolver that is unaware of any available NAT64 device (e.g., probes
configured to use a public DNS resolver instead of the operator
provided resolver). Together, these four tests allow for the detection of a range of NAT64 and
DNS64 setups, including the use of custom prefixes and NAT64 without DNS64.

\section{NAT64 Usage in RIPE Atlas}
\label{sec:nat64-ripe}

We ran the tests described in \S\ref{sec:nat64-search}
on RIPE Atlas with the
goal of understanding use of NAT64 in the wild.
Tests were run over a period of several weeks in December 2022
and January 2023, with 
\NumberOfProbesThatDidThreeTests\ probes participating in the tests.

Table~\ref{tab:total_pass_fail_common_base} summarises the results, showing
the number of probes that consistently pass or fail the tests, and those
that give inconsistent results from repeated runs of the
same test. We observe that most RIPE Atlas probes fail all the tests and
do not access the IPv4 Internet via NAT64, however \ConclusivelyPassedAnyTests\
probes pass one or more of the three primary tests and so may have
NAT64-based access.

\begin{table}
\centering
\caption{Number of probes that passed/failed the DNS tests and the Standard Prefix 
	 ping Test. Note that some probes pass multiple tests.
	 Probes that presented different results to repeated tests are marked
	 ``Inconclusive''.}
\label{tab:total_pass_fail_common_base}
\begin{tabular}{r@{\hskip 2em}r@{\hskip 1em}r@{\hskip 1em}r}
\toprule
  \textbf{Test} &  \textbf{Failed} & \textbf{Passed} & \textbf{Inconclusive} \\
\midrule
  DNS Test 1         &    5938 (96.5\%) &     201 (3.3\%) &            15 (0.2\%) \\
  DNS Test 2         &    6107 (99.2\%) &      42 (0.7\%) &             5 (0.1\%) \\
  Standard Prefix \texttt{ping} test &    6080 (98.8\%) &      66 (1.1\%) &             8 (0.1\%) \\
\bottomrule
\end{tabular}
\vspace{-8mm}
\end{table}

Additionally, \NumberOfProbesThatDidCustomPingTest\ probes either received a non-standard NAT64
prefix for DNS Test 1 or were located in the same
AS as a probe that received a non-standard prefix, and so also
performed the Custom Prefix \texttt{ping} Test.

\pb{Probes with working NAT64 and DNS64}
Probes that pass the DNS tests and have reachability via the NAT64 have the
most functional NAT64 environment of all the probes.
As shown in Table~\ref{tab:total_pass_fail_common_base}, 201 probes pass
DNS Test 1 and 42 probes pass DNS Test 2. Of these, \PassedDNSOneAndTwo\ 
probes passed both DNS tests and can be considered to have fully working
DNS64. 
\SizeNATSixFourPlusDNSSixFour\ of these probes (50\%) consistently passed the appropriate
ping test.
These probes have a fully functional NAT64 and DNS64 setup and comprise
the NAT64+DNS64 group from Figure~\ref{fig:nat64-search-procedure}.
This set contains six IPv6-only and 12 dual-stack probes.

\pb{Probes with working NAT64 but not DNS64} Some probes are not configured with a DNS64 but are able to
reach a NAT64 device: \FailedDNSOneTwoPassedPingTest\ probes failed one or both DNS
tests, but passed one of the ping tests. These probes can reach a NAT64
device, but have, at best, semi-functional DNS64.
120 of these probes were only
able to ping targets via public NAT64 providers, they were excluded from this analysis.
The remaining \SizeNATSixFourOnly\ probes, which were able to ping targets
via a non-public NAT64, comprise the NAT64-only group in Figure~\ref{fig:nat64-search-procedure}.
Of these, \NATSixFourOnlyStdPing\ were able to ping hosts using
the standard NAT64 prefix, while \NATSixFourOnlyCustomPing\ used a custom prefix. 
6 probes could ping via both types of prefixes.

A probe may be categorised as NAT64-only if it is configured to use DNS resolvers without a DNS64 function.
For example, a probe could be explicitly configured to use
a public DNS service rather than its ISP's DNS.
We checked for NAT64-only probes using public resolvers,%
\footnote{\url{https://github.com/trickest/resolvers/blob/main/resolvers-trusted.txt}, plus IPv4 and IPv6 resolvers provided by Google, Cloudflare, Quad9 and OpenDNS}
and found that \UsesPublicResolver\
probes in this set use a public DNS resolver (though they might also use the ISP resolver).
Excluding these probes and the NAT64-only
probes that passed \emph{any} DNS tests, \NoConfiguredNAT\ probes are
likely not intended to use the NAT64. 

\pb{Probes that pass DNS Test 1 without working NAT64}
We found \PassedDNSOneNotPing\ probes that passed DNS Test 1, indicating that they \emph{should}
be able to use NAT64 according to the discovery procedure in RFC
7050~\cite{savolainen_discovery_2013}, but that failed DNS Test 2 and both
ping tests, indicating that they do not
have functioning NAT64.
These probes likely have a misconfigured
DNS resolver (\FreeASOnlyPassedDNSOne\ of the \PassedDNSOneNotPing\ such probes are in AS 12322, ``\emph{Free
SAS}'', suggesting a localised problem). 

We note that probes may pass DNS Test 1 and fail DNS Test 2, but still have
a functioning NAT64. Some DNS64 resolvers only respond to requests for
\texttt{ipv4only.arpa.} to provide the NAT64 prefix and require hosts
to synthesise their own IPv6 addresses \cite{cheshire_special_2020}.

\pb{Summary} Although RIPE Atlas probes that make use of NAT64 are rare, we find
evidence of some usage, with various DNS64 and NAT64 configurations.
We identify \TotalNumNATSixFourProbes\ probes that are potentially behind a NAT64.
\SizeNATSixFourPlusDNSSixFour\ of these have a fully functional NAT64 and DNS64 setup, and 
\SizeNATSixFourOnly\ probes can reach a NAT64 but don't use DNS64,
of which \NoConfiguredNAT\ probes appear to be able to reach the NAT64
by accident. It is not enough to rely on the standard NAT64 prefix discovery, as defined by
RFC 7050~\cite{savolainen_discovery_2013}, and we have shown that our tests
(\S\ref{sec:nat64-search}) provide a more comprehensive approach.

\section{Classifying NAT64 Networks}
\label{sec:probe-categorisation}

\begin{figure}[t]
\centering
\includegraphics[width=0.75\columnwidth,trim={0 12mm 0 8mm}]{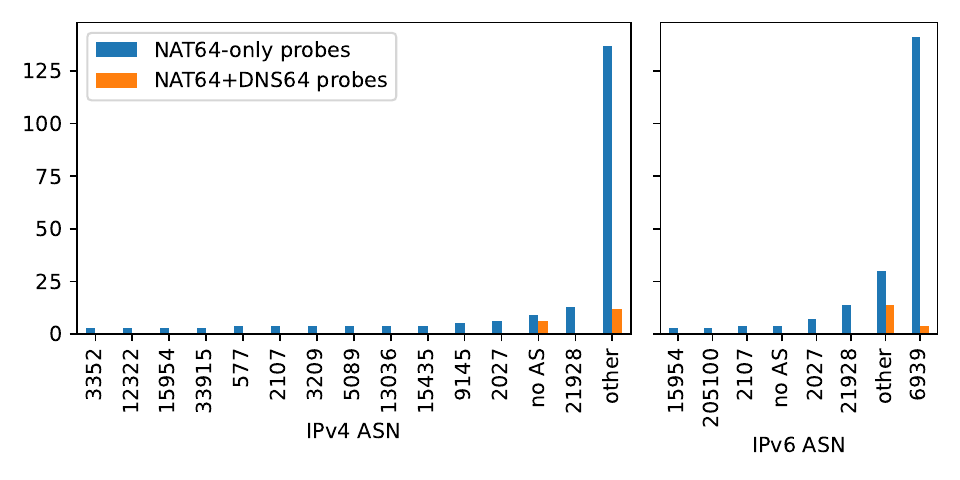}
\caption{ASes of the probes. ``Other'' denotes ASes with two or fewer probes.}
\label{fig:ases_both_groups}
\end{figure}

Probes with reachability to NAT64 middleboxes are located in multiple networks.
Figure~\ref{fig:ases_both_groups} shows the
ASNs of probes using NAT64; we find probes in \TotalNumvSixASN\ IPv6
networks and \TotalNumvFourASN\ IPv4 networks.
Most IPv4 ASes host only one or two NAT64 probes.
The most common IPv6 AS is AS 6939 (Hurricane Electric), hosting 145 probes.
However, as discussed in \S~\ref{sec:excluded}, only one of these
probes is able to traceroute to IPv4 hosts via the NAT64,
suggesting that this is a private NAT64 deployment which blocks outside 
traceroutes.
Only 15 probes are not in any IPv4 AS, so most of these probes can use NAT64 and connect to native IPv4. 

\begin{table}
\centering
\caption{Categorisation of probe IPv6 ASes.}
\label{tab:asn_categorisation}
\begin{tabular}{l@{\hskip 2em}r@{\hskip 1em}r}
\toprule
	\textbf{Category} &  \textbf{ASNs} & \textbf{Probes} \\
\midrule
	Other ISP (business ISP, tier 2 AS, cloud provider, etc) (OI) 
	                              &               7 &               153 \\
 Residential ISP (RI)                        &              18 &                41 \\
	Hobbyist network (run by individuals) (H) &              12 &                14 \\
     Academic and research backbone (A)   &               2 &                 5 \\
    Other (O)                        &               2 &                 4 \\
	Unknown (U)             &               2 &                 3 \\
\bottomrule
\end{tabular}
\end{table}

We use public information (e.g., websites of Internet service providers) to categorise the IPv6 ASes that the probes are in. The results are shown in Table~\ref{tab:asn_categorisation}.
Out of 43 IPv6 ASes, 18 (41.9\%) are residential ISPs; this suggests that NAT64 deployments are uncommon in fixed-line home networks,
and are more common in alternative types of networks.

We use two tests to determine whether NAT64s are provided by the ISP, or are deployed in the local network.
In \textbf{Test 1} we use the DNS test data (\S\ref{sec:nat64-search}) to determine if the default resolver of an AS is a DNS64.
If two probes in an AS query the same DNS64, it is likely provided by the ISP.
Such an ISP likely provides a NAT64 as well.
The DNS tests use the probes' default resolvers, so they expose the configuration of the probes' networks.
A probe can't use a DNS64 ``by accident", while this can happen in the ping tests.

For every NAT64+DNS64 probe, we analysed the results of the DNS tests for all probes in the same AS.
If any other probe used the resolver with which the NAT64+DNS64 probe
passed the DNS tests, we assume that this AS contains an ISP setup.
Two such networks were found: AS 2027 (\emph{MilkyWan}, a residential ISP) and
AS 64475 (\emph{Freifunk Frankfurt}, a public Internet infrastructure project).
This test can over and underestimate the number of ISP setups.
An overestimate can occur if several probes are deployed in a local network with a custom NAT64. 
To mitigate this, we compared the network prefixes of the probes using the DNS64s, which were different.
An underestimate can occur if the AS only contains one probe, if only one probe uses the default ISP resolver, or if there are several ISP resolvers.
3 of the 14 ASes with NAT64+DNS64 probes contain only one probe that participated in the DNS tests, so they could contain undetected ISP deployments. However, eight of the ASes contain 10 or more probes.

Another possible ISP NAT64 setup is in AS 21928 (T-Mobile). Several
probes in this network passed the DNS test some of the time, 
querying link-local addresses which return similar NAT64 prefixes. T-Mobile could be running several DNS64s and NAT64s for redundancy, preventing this setup from being detected.

In \textbf{Test 2}, we use the round-trip time to the NAT64 to detect custom setups. A custom (local) setup will likely have a lower RTT to the NAT64 than an ISP setup.
This test is based on the traceroute data (\S\ref{sec:traceroutes}).
The RTT to the NAT64 is taken to be the RTT to the first hop that begins with the NAT64 prefix.
Looking for probes with an RTT $<$ 2ms to the NAT, we found five probes (2ms was chosen as an estimate of the RTT to a home gateway).
Three of these probes are very likely custom setups (based on public information or contact with the owner).
The fourth probe is in AS 2027, which, based on Test 1, likely has an ISP setup.
We suspect the probe is owned by the network operator and located close to the ISP NAT64, as it is tagged ``Datacentre" and ``Core" on RIPE Atlas. We could not verify whether the fifth probe (ID 11149) is a custom setup. While this test detects some likely home setups, the RTT alone is not a complete indicator of the type of setup.

\begin{figure}[t]
\centering
\includegraphics[width=0.4\columnwidth,trim={0 7mm 0 7mm}]{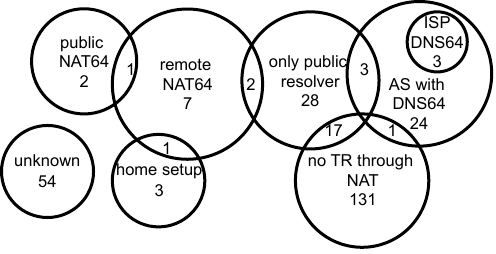}
\caption{Probe categorisation. The size of the circles is not proportional.}
\label{fig:probe-categorisation-euler}
\end{figure}

\pb{Probe categorisation}
To summarise, the probes that were found can be categorised as depicted in
Figure~\ref{fig:probe-categorisation-euler}. Of the \textbf{224} probes:
\begin{itemize}
	\item \textbf{3} likely use an ISP-provided DNS64 (ISP DNS64), while
  \textbf{24} probes are in an AS providing such a DNS64 (AS with DNS64; including
  T-Mobile);
	\item \textbf{3} probes are likely home setups (based on contact with the owner or public information);
	\item \textbf{28} probes in NAT64-only only use a public (non-DNS64) resolver and might be NAT64+DNS64 probes if they used their network's configured resolver;
	\item \textbf{2} probes in NAT64+DNS64 use a public NAT64/DNS64 service;
	\item \textbf{7} probes can traceroute through a NAT64 in another AS
  (remote NAT64);
  	\item \textbf{131} probes can't to traceroute to a NAT64 prefix that they can ping (no TR through NAT);
	\item \textbf{54} probes are uncategorised (unknown).
\end{itemize}

\pb{Summary} NAT64s are present in a variety of IPv4 and IPv6 ASes. Analysing the types of IPv6 ASes, we find that only 41.9\% are domestic ISPs, indicating that NAT64s are more commonly deployed on other types of networks. We find three ASes with a likely ISP-provided NAT64s, and three probes that are likely home setups. Categorising NAT64 deployments is challenging due to a lack of publicly available information.

\section{IPv4 and NAT64 Path Characterisation}
\label{sec:traceroutes}

Having identified RIPE Atlas probes that use NAT64+DNS64 or NAT64-only, we next
perform NAT64 and IPv4 \texttt{traceroute} from the dual stack probes
to known IPv4 targets. We characterise differences in reachability and latency.

\pb{Methodology} Out of \TotalDualStack\ dual-stack RIPE Atlas
probes with NAT64, \TotalNumberProbesRedoTraceroute\ remained online and available
for preliminary measurements (9 NAT64+DNS64 probes, 14 NAT64-only).
We performed traceroute from these probes to 18 IPv4-only targets: seven IPv4-only NTP servers\footnote{We used the IPv4
NTP servers: dodo.mcc.ac.uk, d.st1.ntp.br, time-c-b.nist.gov, ntp1.nog.net.za,
ntp1.st.keio.ac.jp, time-b-g.nist.gov, and ntp2.urz.uni-heidelberg.de}
and 11 RIPE Atlas anchors which were IPv4-only at the time of the
experiment.\footnote{IDs 6771, 6994, 6678, 6827, 6688, 6356, 6366, 6138, 
6712, 6299, and 6711.}
The targets were chosen due to their geographic
spread. The IPv6 address of each target was synthesised
locally using the prefixes obtained from the NAT64 detection tests
(\S\ref{sec:nat64-search}).

\pb{Preliminary traceroutes} Initally we performed single Paris
traceroute~\cite{augustin_avoiding_2006} measurements.
We sent three UDP probe packets for each hop, but to reduce complexity, only
the first detected address for each hop was analysed. 
For every assigned and functional NAT64 prefix on the identified dual stack probes, 
we ran a traceroute to the targets listed above with the aim of
generating one native IPv4 and one NAT64 path for each
probe/prefix/target combination
(most probes can only use a single NAT64 prefix; all probes with multiple NAT64 prefixes are in AS 21928 \emph{T-Mobile}).

\label{sec:excluded}
Our initial result set contains \TotalNumberPathsRedoTraceroute\ pairs of IPv4
and NAT64 paths from the probes to the targets, but a number of paths were
not suitable for analysis.
Of note, \NumberOfPathsNotVisNat\ of the traceroutes to synthesised IPv6 addresses don't
contain any hops starting with the NAT64 prefix. Most of these paths target a NAT64 prefix in AS 6939 (Hurricane
Electric). As these probes were able to ping addresses with this prefix, it is likely that these NAT64 devices
don't respond with ICMPv6 Time Exceeded messages when the hop limit is
reached at the NAT64, and
don't translate ICMPv4 Time Exceeded responses from hosts beyond the NAT64, making the
data for the paths unusable. 
A further 15 pairs of paths were excluded for other anomalies. Excluding these paths, the preliminary measurement set contains
\TotalNumberPathsVisNatTraceroutes\ pairs of paths, starting from 
\TotalNumberProbesVisNatTraceroute\ probes to 18 destinations. 

\pb{Recurring traceroutes} Of the \TotalNumberProbesVisNatTraceroute\
probes with usable results in the preliminary measurements,
\TotalNumberProbesRecurringInitial\ probes were still available when the
second round of measurements was conducted. Traceroutes were performed once
per hour over a span of 41 hours,
to 14 of the 18 targets - four of the Anchor probes used in the preliminary
measurements were disconnected at this time\footnote{One target was no
longer IPv4-only at the time of these measurements, but traceroutes were
performed to its IPv4 address}. Each hour, each probe performed one IPv4 traceroute
per IPv4 target, and one IPv6 traceroute for each combination of translated IPv4 target address and prefix.

Five of these
probes failed to perform traceroutes to all 14 targets during one measurement
run, this data was removed. The 41st measurement run was removed for the
other probes. Six targets did not respond to any traceroutes, traceroutes
to these targets were excluded. Pairs of IPv4 and NAT64 paths from the same probe to the same target, where the NAT64 path did not contain a hop starting with the NAT64 prefix (\TotalNumberPathsRecurringRemoved\ pairs) were removed as well. Thus these measurements resulted in a set of \TotalNumberPathsRecurring\ pairs of IPv4 and NAT64 traceroutes, performed by \TotalNumberProbesRecurring\ probes. Of these probes, \RecurringNATSixFourOnly\ are in the set NAT64-only, \RecurringNATSixFourPlusDNSSixFour\ are NAT64+DNS64 probes. There are \OnePrefixProbesRecurring\ probes that use one NAT64 prefix, 
two probes use two
prefixes,
two probes use three prefixes, and 
three probes use four prefixes.
This measurement set is used as the basis for the following analysis.

\pb{NAT64 locations}
We group paths by the AS that the NAT64 is in, relative to the AS(es) of the
probe. The AS of the NAT64 is taken as the AS of the prefix used (if
non-standard and announced), or otherwise it is taken as the AS of the final
hop in the path that \emph{doesn't} use the NAT64 prefix. If this process yields a different AS for different paths to the same NAT64, the AS corresponding to the hop that is furthest away from a probe is chosen. If the AS can't
be determined, then the NAT64 is assumed to be within the same IPv6 AS as the
probe.

\begin{figure}[t]
\centering
	\includegraphics[width=0.4\columnwidth,trim={0 5mm 0 4mm}]{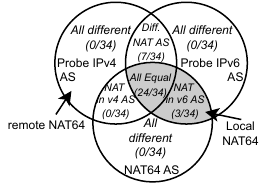}
  \caption{NAT64 and probe locations (shaded: local NAT64, unshaded: remote).}
\label{fig:nat64-locations}
\end{figure}

Figure \ref{fig:nat64-locations} shows the possible NAT64 and probe
locations.
In most cases (24/34), the probe's IPv4 and IPv6 ASes and the AS of the
NAT64 are the same. We consider the
NAT64s in the \emph{all equal} and \emph{NAT in v6 AS} categories to be \emph{local
NAT64s}: they are likely provided by the probe's ISP, or set up in its local network. All other categories are considered \emph{remote
NAT64s}, as the path leaves the
local AS to reach the NAT64. These include public NAT64 services and other NAT64s
that are accessible from outside the AS, e.g. due to a misconfiguration. 

There are \TotalProbesLocalNAT\ probes with a local NAT64 (shaded area in Figure \ref{fig:nat64-locations}), and
seven probes with a remote NAT64 (white area in Figure \ref{fig:nat64-locations}).
Of the probes with a local NAT64, \NumberNATSixFourOnlyLocalNAT/\TotalProbesLocalNAT\
are NAT64-only, compared to 5/7 probes
with a remote NAT. 
One of the five NAT64-only probes with a remote NAT64 can use a NAT64 with
the standard prefix, but is likely not
intended to it: the probe the uses ISP-provided resolver and did not
pass either of the DNS tests. Since the NAT64 is not in its local AS, the
packets are routed to another AS with a NAT64. This AS is likely the
default destination for all packets, and it contains a NAT64 that doesn't
check the source address of packets before translating them. 
This is a minor security concern, as it can be used to hide the source AS of the packets.

\pb{Impact of NAT64 on traceroute}
We first investigate the effect
that NAT64 has on traceroute itself, in terms of
reachability, and the number of missing hops.

The overall success rate (i.e., proportion of traceroutes that reach the target
address, even if it isn't the final hop in the traceroute) across all IPv4
paths is \IPvFourTotalSuccessRate\%, compared to \IPvSixTotalSuccessRate\% for
the paths via NAT64. There are \NumPathsBothSuccessful\
pairs of IPv4 and NAT64 paths from the same probe to the same target that both reached the
destination (\PercentBothPathsSuccessful\% of pairs).
\IPvFourReachedASRate\% of unsuccessful IPv4 paths and \IPvSixReachedASRate\% of unsuccessful NAT64 paths reached the target AS. 
While paths via NAT64 are somewhat less successful than native IPv4 -- with a difference of
\DiffIPvFourIPvSixSuccessRate\ percentage points --  success rates don't differ
much between
groups.

\begin{figure}[t]
\centering
\includegraphics[width=0.6\columnwidth,trim={0 10mm 0 8mm}]{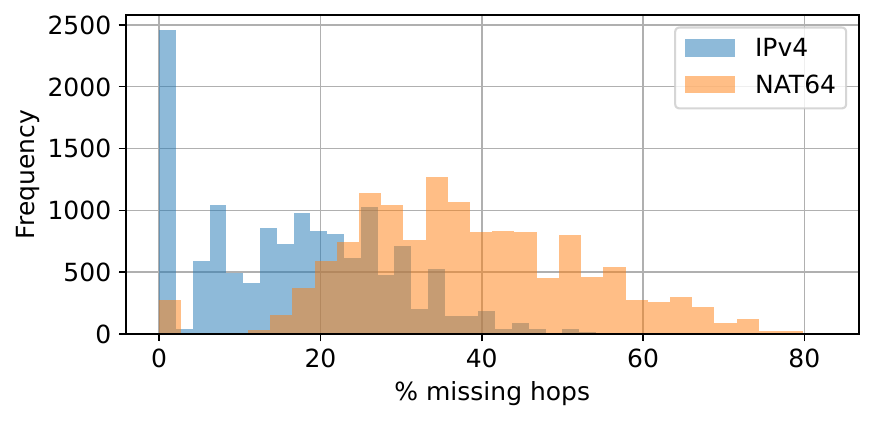}
\caption{Distribution of missing hops in successful pairs of IPv4 and NAT64 paths}
\label{fig:joint-missing-hops-frequency}
\end{figure}

If a hop does not respond to probe packets, then that hop is missing from the
traceroute. Figure \ref{fig:joint-missing-hops-frequency} shows the
distribution of percentages of missing hops in paths where
the IPv4 path and the equivalent path via NAT64 reached the destination.
Paths via the NAT64s have more missing hops: the mean percentage of missing hops for the IPv4 paths is
\AvgMissingHopsIPvFour\% (SD \StdMissingHopsIPvFour, median
\MedianMissingHopsIPvFour\%), via the NAT64 it is
\AvgMissingHopsIPvSix\% (SD \StdMissingHopsIPvSix, median
\MedianMissingHopsIPvSix\%).
A possible explanation for the greater number of missing hops via NAT64 is 
that the NAT64s filter out ICMP response packets. However, this is unlikely:
all of the paths considered here contain at least one hop with a NAT64 prefix.
It is more likely that the missing hops are due to specific routers
filtering ICMP responses on these paths.

To determine how often the same hops are missing in similar traceroutes, 
we compared the NAT64 traceroutes for the two most successful targets in two measurement periods. We
considered all the runs of missing
hops that are preceded and followed by the same
hops in both traceroutes. For example, if traceroute one contains
\texttt{[address 1, missing hop, missing hop, address 2]}, and traceroute
two also contains \texttt{address 1} and \texttt{address 2}, then we check 
if traceroute 2 also contains the exact sequence of hops (i.e., address 1
followed by two missing hops, followed by address 2). 
In the runs of missing
hops identified using this method, on average \MeanSameMissing\% also occurred
in the
second traceroute (median \MedianSameMissing\%, SD
\StdSameMissing\%). This provides 
some evidence that the missing hops are not due to random failure, but
rather caused by specific hops not responding to these traceroutes.

\pb{Impact of NAT64 on path length and latency} 
To analyse NAT64's impact on path length and latency, we consider the \NumPathsBothSuccessful\
pairs of paths where both the IPv4 path and the corresponding
NAT64 path reached the destination.

First, we compare the path length (in IP hops) between IPv4 and NAT64
paths. We consider the path length in hops and not the number of ASes
traversed, because the high number of missing hops in the NAT64 paths makes
it impossible to accurately determine the number of ASes, and because using the number of IP hops provides a finer grained view of possible detours added by using NAT64.

The average path length across
all IPv4 paths is \MeanIPvFourPathLength\ hops (SD \StdDevIPvFourPathLength, median
\MedianIPvFourPathLength), the average NAT64 path length is
\MeanIPvSixPathLength\ hops (SD \StdDevIPvSixPathLength, median
\MedianIPvSixPathLength). As expected, given the potential detours introduced
by NAT64, the NAT64 paths are about three hops longer.
The average length difference for paths with a remote NAT64 is
\RemoteNATMeanPathLenDiff\ hops, for paths with a local NAT64 it is
\LocalNATMeanPathLenDiff. This is caused by the probe in AS 15751, which
appears to have a router on the path that is manipulating
the probing packet's TTL, resulting in traceroutes that reach the destination in very few hops. Excluding this probe, the paths with a remote NAT64 have an average length difference of \RemoteNATMeanPathLenDiffNoOutlier\ hops.
Some of the NAT64 prefixes with the largest length difference are of the form
\texttt{2607:7700:0:\emph{x}:0:\emph{y}}, used by T-Mobile US.
These paths have an average length difference of \XLATAverageLenDiff\ hops; the average
difference of other paths is \NoXLATAverageLenDiff\ hops.

Taking the RTT as measured by traceroute to the first hop 
matching the target, and averaging across all successful UDP probe packets, we find
that the mean RTT across all IPv4 paths is \AvgIPvFourRTT\ ms (SD \StdIPvFourRTT,
median \MedianIPvFourRTT\ ms), the mean NAT64 RTT is \AvgIPvSixRTT\ ms (SD
\StdIPvSixRTT, median \MedianIPvSixRTT\ ms). While the
difference is small, the NAT64 paths have a larger mean RTT. Local NAT64 paths have a higher average RTT
difference (\LocalNATMeanRTTDiff\ ms) than the remote NAT64 paths
(\RemoteNATMeanRTTDiff\ ms), due to the high RTTs of paths using the T-Mobile prefixes (2607:7700:0:x:0:y).
Excluding these the local NAT64 paths have a average RTT difference of
\LocalRTTDiffNoXLAT\ ms.
We also find a moderate correlation between
RTT difference and path length difference (Pearson correlation
coefficient \CorrelationCoefficientLenDiffRTTDiff).

\pb{Summary} NAT64 has a moderate impact on path length and RTT, increasing the
average number of hops by \PercentLenIncrease\% (\AvgPathLenDiff\ hops), and increasing the average
RTT by \PercentRTTIncrease\% (\AvgRTTDiff ms). There is a moderate
correlation between differences in path length and RTT
(Pearson correlation
coefficient \CorrelationCoefficientLenDiffRTTDiff).

\section{Related Work}
\label{sec:related-work}
Parallel work by Hsu et al.~\cite{hsuFirstLookNAT642024} also uses RIPE Atlas to study NAT64 deployments. It differs from our work by not using traceroute, and not focusing on path characteristics such as RTT, number of IP hops, or number of missing hops. Instead, they perform additional ping, DNS and HTTP(S) measurements, and also study the behaviour of public resolvers. They find a similarly low number of NAT64 probes on RIPE Atlas, deployed in a variety of networks\footnote{This work was not yet available to us at the time of writing.}.

Small-scale studies have
evaluated the performance of NAT64 implementations. Lencse and Répás
\cite{lencse_performance_2013} compared the performance of NAT64
implementations in TAYGA and PF under load on a small test network, showing that both degrade gracefully, but that PF has better performance.
Llanto and Yu~\cite{llanto_performance_2012} compare the performance of
NAT44, NAT64 (TAYGA) and native IPv6 on a small test network and compare
NAT44 and NAT64 on a larger university network, showing that while NAT64
and NAT44 had similar performance, the performance of IPv6 was better
than NAT64. Tsetse et al.~\cite{tsetse_measuring_2012} used a small test
deployment to quantify the translation overhead of the IVI translator, a
translator similar to NAT64 used by CERTNET \cite{wu_china_2011}. These studies are very different from the
measurements done in this paper, as they are small-scale, fine grained
measurements of particular NAT64
implementations, performed in a controlled environment. Our work studies the behaviour of NAT64s on the public Internet, and thus does not distinguish between NAT64
implementations. While the number of NAT64s found is relatively small, this work still provides an insight into the prevalence of NAT64, and the behaviour of a number of real-world deployments.

De Vries et al.~\cite{de_vries_how_2015} use RIPE Atlas to explore the difference between the forward and reverse traceroute paths. This is similar
to our study, because it is also a large-scale traceroute measurement study
investigating path similarities.

\section{Conclusions}
\label{sec:conclusions}

We developed tests for identifying NAT64, and applied
them to RIPE Atlas to study NAT64 usage
in the wild. We found that, while RIPE Atlas has around 12,000 probes,
NAT64 usage is rare with only \ConclusivelyPassedAnyTests\ probes able to
use NAT64 to access the IPv4 Internet, and only 18 having a fully functional
NAT64 and DNS64 setup. Importantly, it is not sufficient to
rely on the standard NAT64 prefix discovery procedure
\cite{savolainen_discovery_2013}; our tests
(\S\ref{sec:nat64-search}) are more effective.
Having identified NAT64 use (\S\ref{sec:nat64-ripe}), we
performed \texttt{traceroute} from dual-stack NAT64 probes to compare
IPv4 and NAT64 paths (\S\ref{sec:traceroutes}). On average, the
NAT64 paths were \PercentLenIncrease\% longer, had \PercentRTTIncrease\%
higher RTT, and lower traceroute visibility. NAT64 is a workable substitute
to native IPv4, but impacts latency and reachability.

\balance
\bibliographystyle{splncs04}
\bibliography{nat64-usage}

\appendix
\section{Ethics}

This work does not raise any ethical issues. We identify and characterise the use of NAT64 by probes on a public measurement platform (RIPE Atlas). We generate a relatively small volume of traceroute measurement traffic towards NTP servers, which are publicly listed, and RIPE Atlas anchors, which are architected as measurement targets.

\section{Additional data tables \& figures}

\begin{figure}[t]
\centering
	\includegraphics[scale=0.6]{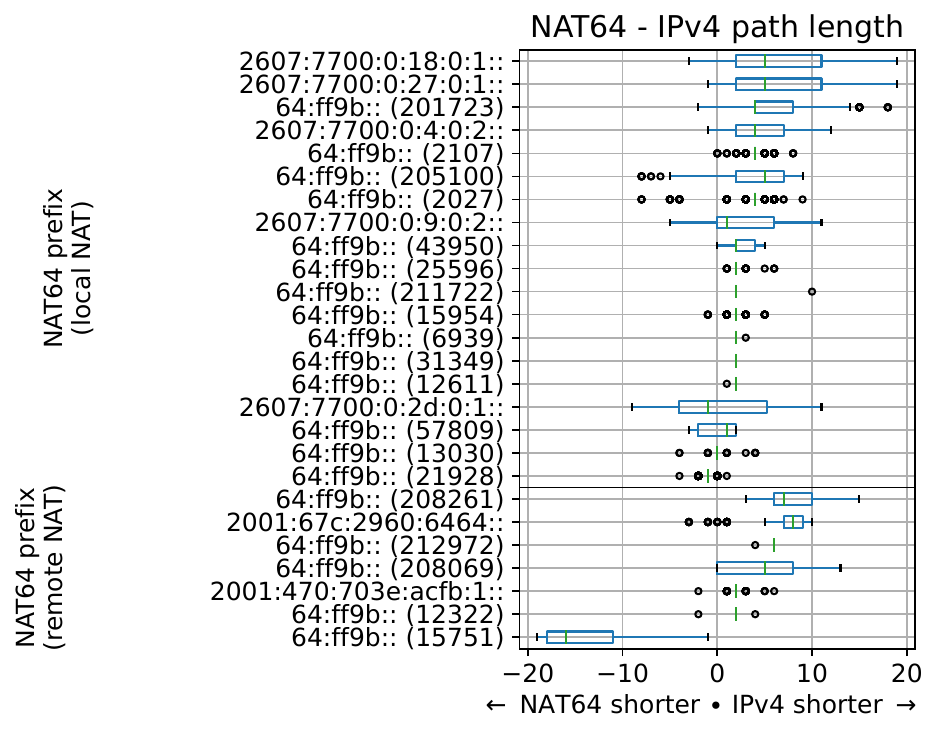}
\caption{Difference in NAT64 and IPv4 path lengths by NAT64 prefix and probe AS.
	}
\label{fig:avg-len-diff-by-nat}
\end{figure}

\begin{figure}[t]
\centering
\includegraphics[scale=0.6]{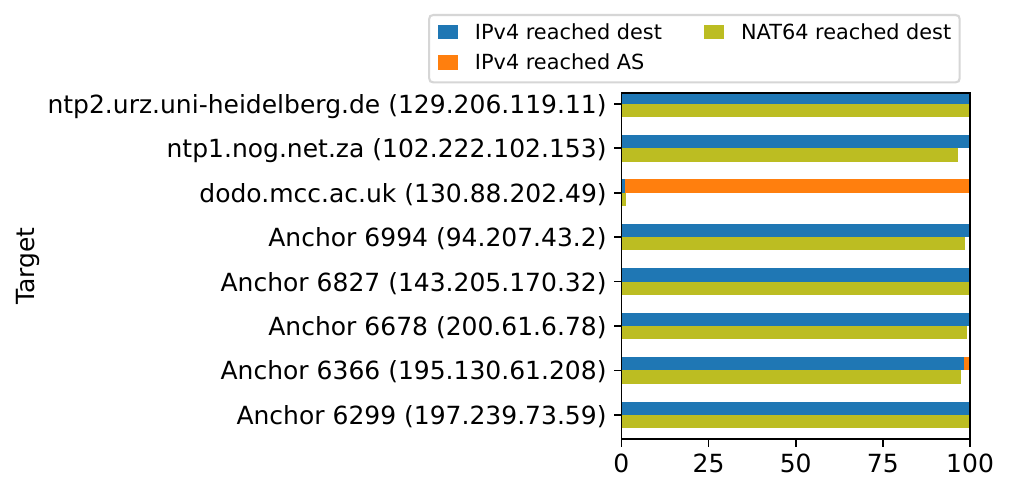}
\caption{\% of paths that reached the destination or destination AS, by target.}
\label{fig:success-rate-by-target}
\end{figure}

\begin{table}
\centering
\caption{Percent success rate for groups of probes/NAT64s.}
\label{tab:success_rate_group_table}
\begin{tabular}{lrr}
\toprule
         \textbf{Set} &  \textbf{IPv4 success} &  \textbf{IPv6 success} \\
\midrule
  NAT64-only &         87.94 &         86.92 \\
 NAT64+DNS64 &         87.92 &         88.07 \\
 Local NAT64 &         88.01 &         86.99 \\
Remote NAT64 &         87.50 &         87.50 \\
\bottomrule
\end{tabular}
\end{table}

\begin{table}
\centering
\caption{ASes of the NAT64-only and NAT64+DNS64 probes. Count of users is based on https://stats.labs.apnic.net/aspop. Some holder names are shortened; category labels are as in Table~\ref{tab:asn_categorisation}.}
	\label{tab:asn_table}
\begin{tabular}{rllrr}
\toprule
	\textbf{ASN} &                                        \textbf{Holder name} & \textbf{Category} &  \textbf{Users (est.)} &  \textbf{\# probes} \\
\midrule
  6939 &                                          Hurricane Electric &       OI &        793668 &       145 \\
 21928 &                                   T-Mobile US &       RI &      19377903 &        14 \\
  2027 &                    MilkyWan Association &       RI &           479 &         9 \\
  2107 &                                  ARNES-NET - ARNES &        A &         11923 &         4 \\
205100 &                            F3 Netze e.V. &        O &          4148 &         3 \\
 15954 &  Tecnocratica Centro de Datos, S.L. &       OI &           --- &         3 \\
133481 &                          AIS Fibre &       RI &       5598723 &         2 \\
 15751 & Meteor Mobile Communications Limited &       RI &        286273 &         2 \\
203528 &     Fabrizzio Jose Petrucci &        H &           --- &         2 \\
201723 &                 R. van der Meijden &        H &           --- &         2 \\
 43950 & SPILSBY-AS - Terence Froy trading as "Spilsby I... &        U &           --- &         2 \\
  7713 &        PT Telekomuni\-kasi Indonesia &       RI &      31773744 &         1 \\
  3320 &                         Deutsche Telekom AG &       RI &      24767366 &         1 \\
  3209 &                            Vodafone GmbH &       RI &      17566119 &         1 \\
 12322 &                                  Free SAS &       RI &       7223746 &         1 \\
  1136 &                                     KPN B.V. &       RI &       5763278 &         1 \\
 45629 & JasTel Network International Gateway &       RI &       5404250 &         1 \\
  2527 &            Sony Network Communications &       RI &       3925106 &         1 \\
  6568 &                              EntelNet &       RI &       2594067 &         1 \\
 34779 &                               T-2, d.o.o. &       RI &        266187 &         1 \\
  1916 &                 Rede Nacional de Ensino e Pesquisa &        A &         68162 &         1 \\
  4800 &          LINTASARTA-AS-AP PT Aplikanusa Lintasarta &       OI &         42633 &         1 \\
 25596 &            Cambrium IT Services B.V. &       OI &         27970 &         1 \\
 12611 & RKOM &       RI &         24572 &         1 \\
 57809 &                              SERVEURCOM - UNYC SAS &       OI &         19850 &         1 \\
 13030 &                   Init7 (Switzerland) Ltd. &       RI &         15863 &         1 \\
 31349 &                       a-net Liberec s.r.o. &       RI &         15719 &         1 \\
206238 &              Freedom Inter\-net BV &       RI &         13410 &         1 \\
 25660 &                                                CTC &        U &          1698 &         1 \\
 29670 &      Individual Network Berlin e.V. &       RI &           --- &         1 \\
211579 &                          Laura Hausmann &        H &           --- &         1 \\
212037 &                      Marc Provost &        H &           --- &         1 \\
203062 &                    Alexandre Roux &        H &           --- &         1 \\
213204 &                             Marvin Gaube &        H &           --- &         1 \\
208069 &                            Cecile Morange &        H &           --- &         1 \\
213318 &                                               None &        H &           --- &         1 \\
208261 &                     Pomme Telecom SASU &        H &           --- &         1 \\
212972 &          Nicolas VUILLERMET &        H &           --- &         1 \\
 62538 &                                             IPFAIL &        H &           --- &         1 \\
204345 &                        Timothy Stallard &        H &           --- &         1 \\
 64475 & Freifunk Frankfurt am Main e.V. &        O &           --- &         1 \\
 56381 & level66.network UG &       OI &           --- &         1 \\
211722 &                        Nullroute ry &       OI &           --- &         1 \\
\bottomrule
\end{tabular}
\end{table}

\end{document}